\documentclass[preprint,nofootinbib]{revtex4}
\usepackage{graphicx}
\usepackage{amsmath,amssymb}

\newcommand{\be}{\begin{equation}}
\newcommand{\ee}{\end{equation}}
\newcommand{\bea}{\begin{eqnarray}}
\newcommand{\eea}{\end{eqnarray}}

\newcommand{\cl}{{\rm Cl}_2}
\newcommand{\dl}{{\rm Li}_2}
\newcommand{\tl}{{\rm Li}_3}

\newcommand{\nd}{{\mathbf n}}

\newcommand\bef{\begin{figure}}
\newcommand\eef[1]{\label{fg:#1}\end{figure}}
\newcommand\beq{\begin{equation}}
\newcommand\eeq[1]{\label{#1}\end{equation}}
\newcommand\beqa{\begin{eqnarray}}
\newcommand\eeqa[1]{\label{#1}\end{eqnarray}}
\newcommand\bet{\begin{table}}
\newcommand\eet[1]{\label{tb:#1}\end{table}}

\newcommand\fgn[1]{Figure \ref{fg:#1}}
\newcommand\eqn[1]{Eq.\ (\ref{#1})}
\newcommand\scn[1]{Section \ref{sec:#1}}

\begin{document}

\date{\today}

\title{
The relevance of the three dimensional Thirring coupling at finite temperature and density\footnote{\it A joyous exercise using polylogarithms and related special functions performed in isolation.}
}
\author{R. Narayanan}
\affiliation{Department of Physics, Florida International University, Miami,
FL 33199.}

\begin{abstract}
We studied the three dimensional Thirring model in the limit of infinite number of flavors at finite temperature and density. We calculated the number density
as a function of temperature and the density at zero temperature serves as a relevant parameter. A three dimensional free fermion gas behavior as the
density at zero temperature approaches zero smoothly crosses over to a 
two dimensional free fermion gas behavior as the density at zero temperature approaches infinity.
\end{abstract}

\maketitle 

\section{Introduction and Summary}

The three dimensional Euclidean (two spatial and one thermal) Thirring model with $N$ flavors of two-component fermions would have been deemed non-renormalizable by a standard power counting argument but it has been shown to be renormalizable in a $\frac{1}{N}$ expansion~\cite{Parisi:1975im,Hikami:1976at,Yang:1990ki,Gomes:1990ed,Hands:1994kb}. Recently, this strongly coupled theory has been extensively studied to explore the possibility of mass generation~\cite{DelDebbio:1995zc,Hands:1996px,DelDebbio:1997dv,DelDebbio:1999he,Hands:1999id,Christofi:2007ez,Christofi:2007ye,Gies:2010st,Janssen:2012pq,Schmidt:2015fps,Wellegehausen:2017goy,Hands:2017hhk,Hands:2018vrd,Hands:2018kvr,Lenz:2019qwu}.
With the possible exception of the $N=1$ model, a spontaneous generation of mass is most likely ruled out. A numerical analysis of QED in three dimensions has also resulted in similar observations~\cite{Karthik:2015sgq,Karthik:2016ppr,Karthik:2017hol,Karthik:2016ixj,Karthik:2018tnh}
but monopoles present in QED can become relevant~\cite{Karthik:2019mrr}. 

In this paper, we explore the relevance of the Thirring coupling in the large $N$ limit at finite temperature and density. We consider only the massless theory owing to the previous analysis of spontaneous mass generation. The physics at zero temperature has been briefly sketched out in~\cite{Goykhman:2016zgd}. The effective action in the large $N$ limit after introducing the standard vector auxiliary field is complex for a non-zero chemical potential and this leads us to perform a saddle point analysis. There are several saddle points at a fixed chemical potential and temperature but we will provide a graphical proof that only one particular saddle point dominates all all values of chemical potential and temperature.

Let $T$ and $\mu$ stand for the temperature and chemical potential measured in units of inverse of the Thirring coupling per flavor, $\lambda$.
We will show that the number density at zero temperature in units of $\lambda$ is
\be
\nd_0 =  \mu + 2\pi\left ( 1- \sqrt{1+\frac{\mu}{\pi}} \right),
\ee
and can be used to set the chemical potential. The relevance of the Thirring coupling is already seen by noticing that the number density at zero temperature  smoothly crosses over from
$\frac{\mu^2}{4\pi}$ as $\mu\to 0$ to $\mu$ as $\mu\to\infty$. 
Defining a reduced temperature by $T=\sqrt{4\pi \nd_0} t$
and writing the number density as a function of temperature as $\bar\nd(\nd_0,t)\nd_0$,
we will show that $\bar\nd(\nd_0,t)$ is the solution to
\be
\bar\nd(\nd_0,t) =8 t^2 r_2\left ( \frac{1 + \sqrt{\frac{\nd_0}{4\pi}}[1-\bar\nd(\nd_0,t)]}{2t} , 0\right),\qquad 1 \le \bar\nd(\nd_0,t) \le 1+\sqrt{\frac{4\pi}{\nd_0}};
\ee
where
\be
 r_2(u,\theta) =  \frac{\pi^2}{24} + \frac{u^2}{2} -\frac{\theta^2}{2}+\sum_{k=1}^\infty (-1)^{k} \frac{ \left(e^{-2ku}\right)\cos(2k\theta)}{2k^2};
\qquad u>0;\quad -\frac{\pi}{2} \le \theta < \frac{\pi}{2}.
\ee
This result for the number density is plotted as a function of temperature in \fgn{numden} and shows that it smoothly crosses over from a three dimensional
free fermion gas as $\nd_0 \to 0$ to a two dimensional free fermion gas as $\nd_0 \to \infty$. 
\bef
\centering
\includegraphics[scale=0.65]{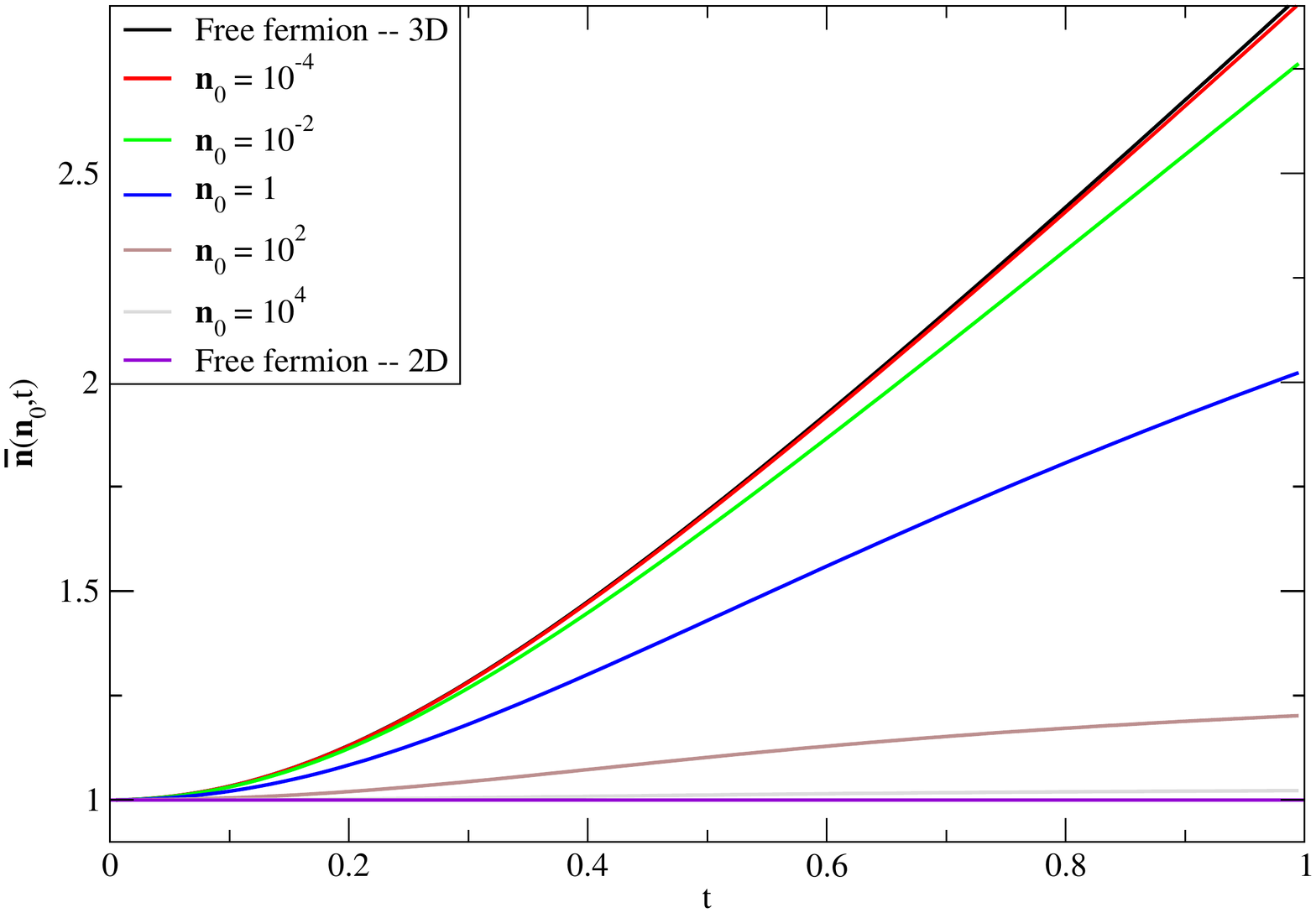}
\caption{ Number density as a function of temperature for different choices of density at zero temperature.
}
\eef{numden}

The rest of the paper is organized as follows. We set up our notation for the three dimensional Thirring model and arrive at the saddle point equations
in the limit of $N\to\infty$ in \scn{thirring}. The saddle point that dominates at all chemical potential and temperature is analyzed 
in \scn{sad0} to obtain the main results stated above. Due to the involved inter-dependencies of the different saddle points at a fixed temperature
and chemical potential, we revert to a graphical analysis in \scn{sadne0} to show that the saddle point discussed in \scn{sad0} dominates at
all chemical potential~\footnote{\sl Much ado about nothing -- William Shakespeare.}.

\section{The three dimensional continuum Thirring model}\label{sec:thirring}

The action for the continuum Thirring model in three Euclidean dimensions is given by
\be
S(\bar\psi_i,\psi_i;\lambda) = \int d^3 x \sum_{i=1}^N \bar \psi_i  D(\mu) \psi _i 
+ \frac{1}{2N\lambda} \int d^3 x \sum_{k=1}^3 \left( \sum_{i=1}^N \bar\psi_i \sigma_k \psi_i\right)^2,\label{thirring}
\ee
where
\be
D(\mu) = \sum_{k=1}^3 \sigma_k \partial_k +\sigma_3 \mu
\ee
is the Dirac operator acting on a two-component fermion
in a $\ell^2\times \beta$ periodic box and $\mu$ is the chemical potential.
The fermions obey periodic boundary conditions in the spatial directions and anti-periodic boundary conditions in the thermal direction.
Upon introduction of a vector auxiliary field, $V_k(x)$, to replace the four-fermi interaction and  a subsequent integration of the fermions
results in
\be
S(V_k,\lambda) = N \left[ \frac{\lambda}{2}\left(\int d^3 x   \sum_k V_k^2\right) - \ln \det (D(V_k)) \right] ;
\ee
where
\be
D(V_k) = \sum_{k=1}^2 \sigma_k ( \partial_k +iV_k) + \sigma_3 \mu
\ee
With $V_k \to -V_k$ and $x_k \to -x_k$, we see that $\mu \to -\mu$. Therefore, it is sufficient to consider a positive value for the chemical
potential in our analysis.

Assuming translational invariance to hold in the large $N$ limit, we will analyze the 
action per flavor with the auxiliary field restricted to constants,
\be 
V_1(x) = \frac{2\pi h_1}{\ell};\qquad V_2(x) = \frac{2\pi h_2}{\ell} ;\qquad V_3(x) = \frac{2\pi h_3}{\beta}.
\ee
The minimum will occur at $h_1=h_2=0$ in the $\ell\to\infty$ limit. One can use standard formulas in~\cite{GradRyz:2007} to perform the sum over
momentum in the $\beta$ direction and the resulting action density (per unit spatial volume) per flavor is
\be
S(h_3;T,\mu) = 2\pi^2 T h_3^2 
-\frac{2T^2}{\pi}  \int_0^\infty dx\ x  \ln \frac{ 4\cosh\left(x+i\pi h_3 + \frac{\mu}{2T}\right)
\cosh\left(x-i\pi h_3 - \frac{\mu}{2T}\right)}{e^{2x}} 
\label{faction}
\ee
The integration variable is related to the spatial momentum by $x=\frac{\beta p}{2} = \frac{p}{2T}$ and
we have set $T$ and $\mu$ in units of $\lambda$.

The action density is complex and we will perform a saddle point analysis.
We elevate $(2\pi h_3)$ to a complex variable $z = \gamma +i\frac{\delta}{T}$ and define
\be
 \omega = u+i\theta;\quad u=\frac{\mu-\delta}{2T};\quad \theta\in \left[ \frac{-\pi}{2},\frac{\pi}{2}\right];\qquad
z=-2i\left(\omega - \frac{\mu}{2T} + in\pi\right);\quad n\in \mathbf{I}.
\ee
The action density can be written in terms of  polylogarithms~\cite{NIST:DLMF} as
\be
S(n,\omega;T,\mu) = S^b(n,\omega;T,\mu) + S^f(\omega;T,\mu);
\ee
\be
S^b(n,\omega;T,\mu) = -2T \left(\omega-\frac{\mu}{2T} + in\pi\right)^2;\qquad S^f(\omega;T,\mu)= \frac{T^2}{2\pi} \left [ \tl\left(-e^{2\omega}\right) + \tl\left(-e^{-2\omega}\right) \right].\label{actden}
\ee
The partition function is given by
\be
Z(T,\mu) = \sum_{n=-\infty}^\infty \int_{-\frac{i\pi}{2}}^{\frac{i\pi}{2}} d\omega e^{ -N \ell^2 S(n,\omega;T,\mu)}.
\ee
The saddle points are given by $\frac{dS(n,\omega;T,\mu)}{d\omega}=0$
which we will label by $\omega_n$ or $(\theta_n,\delta_n)$ with the possibility that more than one saddle point exist at a fixed $n$.

The number density per flavor is given by
\bea
\nd(T,\mu) &=& \frac{T}{\ell^2 N} \frac{d \ln Z(T,\mu)}{d\mu} 
= -\frac{1}{2Z(T,\mu)} \sum_{n=-\infty}^\infty \int_{-\frac{i\pi}{2}}^{\frac{i\pi}{2}} d\omega\frac{dS_f(n,\omega;T,\mu)}{d\omega}e^{ -N \ell^2 S(n,\omega;T,\mu) }\cr
&=& -\frac{1}{2Z(T,\mu)} \sum_{n=-\infty}^\infty \int_{-\frac{i\pi}{2}}^{\frac{i\pi}{2}} d\omega \left( 4T\left(\omega-\frac{\mu}{2T} + in\pi\right) + \frac{dS(n,\omega;T,\mu)}{d\omega}\right)e^{-N \ell^2 S(z;T,\mu)}.\cr
&&
\label{numden}
\eea

\subsection{Equations for saddle points}\label{sec:saddle}

The complex valued equation for the saddle point using polylogarithm identities is
\bea
\frac{\mu}{2T} &=& \omega_n +in\pi +\frac{T}{4\pi} \left [ \dl\left(-e^{-2\omega_n}\right) - \dl\left(-e^{2\omega_n}\right)\right]\cr
&=&\begin{cases}
\frac{   \pi T}{24} + \omega_n +in\pi +\frac{ T}{2\pi} \omega_n^2 +\frac{T}{2\pi} \sum_{k=1}^\infty (-1)^k \frac{e^{-2k \omega_n}}{k^2};  & u_n > 0 \cr
 -\frac{   \pi T}{24} + \omega_n +in\pi -\frac{ T}{2\pi} \omega_n^2 -\frac{T}{2\pi} \sum_{k=1}^\infty (-1)^k \frac{e^{2k \omega_n}}{k^2}; & u_n < 0
\end{cases}.
\eea

For the purpose of analysis,
we separate the above equation into its
real and imaginary part as
\bea
\frac{\pi (\theta_n+n\pi)}{T} &=& r_1(|u_n|,\theta_n) = -|u_n|\theta_n+  \sum_{k=1}^\infty (-1)^k \frac{ e^{-2k|u_n|}\sin(2k\theta_n)}{2k^2};\cr
\frac{\pi \mu}{2T^2} &=& \frac{u_n}{|u_n|}\left[ \frac{\pi |u_n|}{T} + r_2(|u_n|,\theta_n)\right]\cr
 r_2(|u_n|,\theta_n) &=&  \frac{\pi^2}{24} + \frac{u_n^2}{2} -\frac{\theta_n^2}{2}+\sum_{k=1}^\infty (-1)^{k} \frac{ \left(e^{-2k|u_n|}\right)\cos(2k\theta_n)}{2k^2}.
\label{sadeqn}
\eea
The real and imaginary parts of the action density needed to study the dominance of one saddle point over another are
\bea
S_R^b(n,\theta,\delta;T,\mu) &=& 2T(\theta+n\pi)^2 -\frac{\delta^2}{2T} \cr
S_R^f(\theta,\delta;T,\mu) &=&\frac{2T^2}{\pi} \left[ |u| \left( \theta^2 -\frac{\pi^2}{12}\right) -\frac{|u|^3}{3}+\sum_{k=1}^\infty
(-1)^k \frac{e^{-2k|u|} \cos(2k\theta)}{2k^3}\right] \cr
S_I^b(n,\theta,\delta;T,\mu) &=& 2\delta(\theta+n\pi) \cr
S_I^f(\theta,\delta;T,\mu)&=&-\frac{u}{|u|} \frac{2T^2}{\pi}\left [   \frac{\theta^3}{3} - \frac{\pi^2\theta}{12} 
+ u^2\theta +\sum_{k=1}^\infty
(-1)^k \frac{e^{-2k|u|} \sin(2k\theta)}{2k^3}
\right].\label{actdenri}
\eea

Referring to \eqn{derr12ut}, we note that the derivative of $r_1(|u|,\theta)$ with respect to $\theta$ at $\theta=0$ is negative. The derivative can become zero at most once for $|\theta| < \frac{\pi}{2}$, if $|u| < u_0 = \frac{1}{2}\ln \frac{3+\sqrt{5}}{2}$.
Referring to \eqn{ku1}, we conclude that $r_1(|u|,\theta)$ is
positive for $-\frac{\pi}{2} \le \theta < 0$ and negative for $0 < \theta \le \frac{\pi}{2} $. There is always a saddle point with $n=0$ and it has $\theta_0=0$.
For $n \ne 0$, the saddle point, if one exists, will be in the region $-\frac{\pi}{2} \le \theta_n < 0$ if $n >0$ and 
will be in the region $0 < \theta_n \le \frac{\pi}{2} $ if $n<0$.

Assuming the saddle point at $n=0$ dominates, the result for the number density will be
\be
\nd(T,\mu) = \delta_0.
\ee
Let us assume we have a solution to the saddle point equations given by $(\theta_n,\delta_n)$ with $n>0$. 
Since
\be
r_1(|u_n|,-\theta_n) = -r_1(|u_n|,\theta_n);\qquad r_2(|u_n|,-\theta_n) = r_2(|u_n|,\theta_n),\qquad u_{-n} = u_n
\ee
we can conclude that $(\theta_{-n},\delta_{-n}) = (-\theta_n,\delta_n)$ is a saddle point with $-n < 0$.
It follows from \eqn{actdenri} that
\be
S_R(n,\theta_n,\delta_n;T,\mu) = S_R(-n,\theta_{-n},\delta_{-n};T,\mu);\quad
S_I(n,\theta_n,\delta_n;T,\mu) = -S_I(-n,\theta_{-n},\delta_{-n};T,\mu).
\ee
If this pair of saddle points were to dominate over the solitary one at $n=0$, the number density will be
\be
\nd(T,\mu) = \delta_n - 2(\theta_n+n\pi )T \tan [N\ell^2 S_I(n,\theta_n,\delta_n;T,\mu)]
\ee
which does not have a smooth $N\to\infty$ limit.

\section{Analysis assuming the saddle point at $n=0$ dominates}\label{sec:sad0}

The only equation is
\be
\frac{\pi \mu}{2T^2} =   \frac{u_0}{|u_0|} \left [ \frac{\pi |u_0|}{T} + r_2(|u_0|,0)\right].
\ee
Referring to \eqn{ku2}, we have $r_2(|u_0|,0) > 0$ which results in
\be
u_0 > 0; \quad \delta_0 = \frac{2T^2}{\pi} r_2\left(\frac{\mu-\delta_0}{2T},0\right).
\label{sadneq0}
\ee
Rewriting the saddle point equation as
\be
 \frac{\mu}{T^2} -\frac{2u_0}{T} = \frac{2}{\pi} r_2(u_0,0);\qquad u_0=\frac{\mu-\delta_0}{2T} > 0.\label{sadgeq0}
\ee
we see there is  one and only one solution for $u_0$  such that $0 \le u_0 \le \frac{\mu}{2T}$ ($ 0 \le \delta_0 \le \mu$).
Using \eqn{derr12ut},we obtain
\be
\frac{\partial \delta_0}{\partial T} = \frac{2T}{\pi} \frac{ 2r_2(u_0,0) + u_0 d_1(u_0,0)}{1-\frac{T}{\pi} d_1(u_0,0)}.
\ee
We note from \eqn{ku2} and \eqn{derr12ut} that
\be
d_1(u_0,0)  < 0;\qquad 
2r_2(u_0,0) + u_0 d_1(u_0,0) = u_0 \ln (1+e^{-2u_0}) +4\int_0^{u_0} dx \frac{x}{1+e^{2x}} > 0.
\ee
Therefore
we conclude that $\frac{\partial \delta_0}{\partial T} > 0$ and $\delta_0$ is a monotonically increasing function of temperature
at a fixed chemical potential.

Assuming that the saddle point at $n=0$ dominates at all chemical potential and temperature, the number density is given by $\nd=\delta_0$.
Since
\be
\lim_{T\to 0} \frac{2T^2}{\pi} r_2(u_0,0) = \frac{(\mu-\delta_0)^2}{4\pi}.
\ee
the saddle point equation as $T\to 0$ is
\be
\nd_0 =  \frac{(\mu-\nd_0)^2}{4\pi},
\ee
which results in the number density at $T=0$ to be
\be
\nd_0 =  \mu + 2\pi\left ( 1- \sqrt{1+\frac{\mu}{\pi}} \right).
\ee
We can trade the chemical potential for $\nd_0$ using
\be
\mu = \nd_0 +\sqrt{4\pi \nd_0}.\label{thchden}
\ee

To see the relevance of the Thirring coupling, we define a reduced temperature by $T=\sqrt{4\pi \nd_0} t$ and write the number density at any temperature as
$[ \bar\nd(\nd_0,t) \nd_0]$,
noting that $\bar\nd$ will depend on $\nd_0$ in addition to $t$. It is this dependence that shows the relevance of the Thirring coupling.
Referring to \eqn{sadneq0}  and \eqn{thchden} we arrive at
\be
\bar\nd(\nd_0,t) =8 t^2 r_2\left ( \frac{1 + \sqrt{\frac{\nd_0}{4\pi}}[1-\bar\nd(\nd_0,t)]}{2t} , 0\right).
\ee
One can either use \eqn{sadeqn} or \eqn{ku2} and see that the above equation reduces to a free fermion behavior in three dimensions as $\nd_0\to 0$.
Using \eqn{ku2} and \eqn{sadgeq0}, the asymptotic behavior in $t$ at  a fixed $\nd_0$ is
\be
\bar\nd(\nd_0,t) = 1 + \sqrt{\frac{4\pi}{\nd_0}} - \frac{\pi \left(1 + \sqrt{\frac{4\pi}{\nd_0}}\right)}{(\ln 2)\sqrt{4\pi \nd_0}} \frac{1}{t} +O(t^{-2}).
\ee
Since the number density monotonically increases with temperature at a fixed $\nd_0$, it is bounded by
\be
1 \le \bar\nd(\nd_0,t) \le 1 + \sqrt{\frac{4\pi}{|\nd_0|}}
\ee
The density does not change with temperature in the limit $\nd_0\to\infty$ and this is the free fermion behavior in two dimensions.
The sub-leading term for small $t$ is
\be
\bar\nd(\nd_0,t) = 1 + \frac{\pi^2}{3} \frac{1}{1+\sqrt{\frac{\nd_0}{\pi}}} t^2 + O(t^3).
\ee
A plot of $\bar\nd(\nd_0,t)$ as a function of $t$ has already been shown in \fgn{numden} at several different values of $\nd_0$ to demonstrate the crossover from three dimensional free fermion behavior as
$\nd_0\to 0$ to two dimensional free fermion behavior as $\nd_0\to\infty$.

\section{Non-dominance of the saddle points at $n\ne 0$}\label{sec:sadne0}

We will only consider $n>0$ since we have shown at the end of \scn{saddle}
that saddle points at $n$ and $(-n)$ are paired. Then every solution has to satisfy $-\frac{\pi}{2} \le \theta_n \le 0$.
We will fix the temperature and  use $\theta_n$  instead of chemical potential since it is more convenient from the viewpoint of solving the saddle point equations.
We will  graphically demonstrate the non-dominance of the saddle points at $n\ne 0$ by defining
\be
\Delta(n,\theta_n,T) = \frac{ S_R(n,\theta_n,\delta_n(\theta_n);\mu(\theta_n),T) - S_R(0,0,\delta_0(\theta_n);\mu(\theta_n),T)}{2T n^2 \pi^2} 
\ee
and showing that it remains positive at all $n$, $T$ and allowed values of $\theta_n \in \left[ -\frac{\pi}{2}, 0\right]$. To this end we will use the following steps:
\begin{enumerate}
\item We will show that $\mu\to \infty$ as $\theta_n \to 0$ at all temperatures.
\item There exists a temperature $T_0(n)$ above which a certain region given by  $-\frac{\pi}{2} < \theta_l(T) < \theta_n < \theta_r(T) < 0$ has
no solution to the saddle point equations. The chemical potential will monotonically increase from $0$ to $\infty$ as $\theta_n$ increases from $\theta_r(T)$ to $0$
and it will monotonically increase from $0$ to a finite non-zero value as $\theta_n$ decreases from $\theta_l(T)$ to $-\frac{\pi}{2}$.
Furthermore, $\theta_l(T_0(n)) = \theta_r(T_0(n)) = \theta_0(n)$.
\item There exists a temperature $T_1(n)$ below which the chemical potential will monotonically decrease from $\infty$ to a finite non-zero value as $\theta_n$
decreases from $0$ to $-\frac{\pi}{2}$. In other words there will be a region of chemical potential given by $ 0 \le \mu < \mu_1(T)$ for which there is no
solution to the saddle point equations. Furthermore, $\mu_1(T_1(n))=0$.
\item We will explicitly study the case of zero temperature and the case of zero chemical potential.
\item Even though $\Delta(n,\theta_n,T)$ will not be monotonic in $\theta_n$ at a fixed $n$ and temperature, it reaches an absolute minimum at $\theta=-\frac{\pi}{2}$
at all $n$ and temperatures. We will study $\Delta\left(n,-\frac{\pi}{2},T\right)$ as a function of $T$ and show that the minimum at each $n$ is larger than zero for all $n$.
\end{enumerate}
\subsection{$\theta_n \to 0_- \Rightarrow \mu\to \infty$:}
 Keeping terms to relevant orders in $\theta_n$ and
using \eqn{derr12ut}, the first equation in \eqn{sadeqn} results in 
\be
|u_n| = -\frac{n\pi^2 }{T \theta_n} -\frac{\pi}{T} +O\left(e^{\frac{1}{\theta_n}}\right).
\ee
Inserting this into the second equation in \eqn{sadeqn} results in
\be
 u_n > 0;\quad \mu = \frac{n^2\pi^3}{\theta_n^2} + \frac{\pi T^2}{12} - \pi + O(\theta_n^2).
\ee
If we use the diverging chemical potential with the leading correction into \eqn{sadgeq0}, we find that $u_0 = u_n$ up to the two correction terms.
Therefore, we arrive at
\be
\Delta(n,0_-,T)
= 1,
\ee
and the saddle point at $n=0$ dominates as $\mu\to\infty$ for all values of temperature.

\subsection{$T_0(n)$:}
Noting that
$d_2(|u|,\theta) > 0$ for $-\frac{\pi}{2} < \theta < 0$, the right hand side of the first equation in \eqn{sadeqn} monotonically increases with $|u|$ at a fixed $\theta$. Therefore,
at every value of $\theta$, the lowest value of the right hand side is given by $r_1(0,\theta)$. If the temperature is above a certain value,
we will have a region of $\theta$ where there is no solution to the first equation in \eqn{sadeqn}. This transition temperature is the
solution to
\be
\frac{\pi}{T_0(n)} \left (\theta_0(n) + n\pi\right) = r_1(0,\theta_0(n));\qquad \frac{\pi}{T_0(n)} = -\ln [2\cos\theta_0(n)],
\label{sadeqnt0}
\ee
which results in
\bea
\theta_0(1) = -0.372690809\ \pi;& \theta_0(2) = -0.350144201\ \pi;& \theta_0(5) = -0.339555189\ \pi\cr
T_0(1) = 12.563174152; & T_0(2)= 2\times 16.123729198;& T_0(5) = 5\times 18.139848147;
\eea
and
\be
 \theta_0(\infty) = \frac{\pi}{3};\qquad \lim_{n\to\infty} \frac{T_0(n)}{n} = \frac{\pi^2}{r_1\left(0,\frac{\pi}{3}\right)} = 19.448615248.
\ee
Since $u_n=0$ at this temperature, $T_0(n)$,  and $\theta_0(n)$ we have $r_2(0,\theta_0(n))=0$ and $\mu=0$ at this point.

\bef
\centering
\includegraphics[scale=0.625]{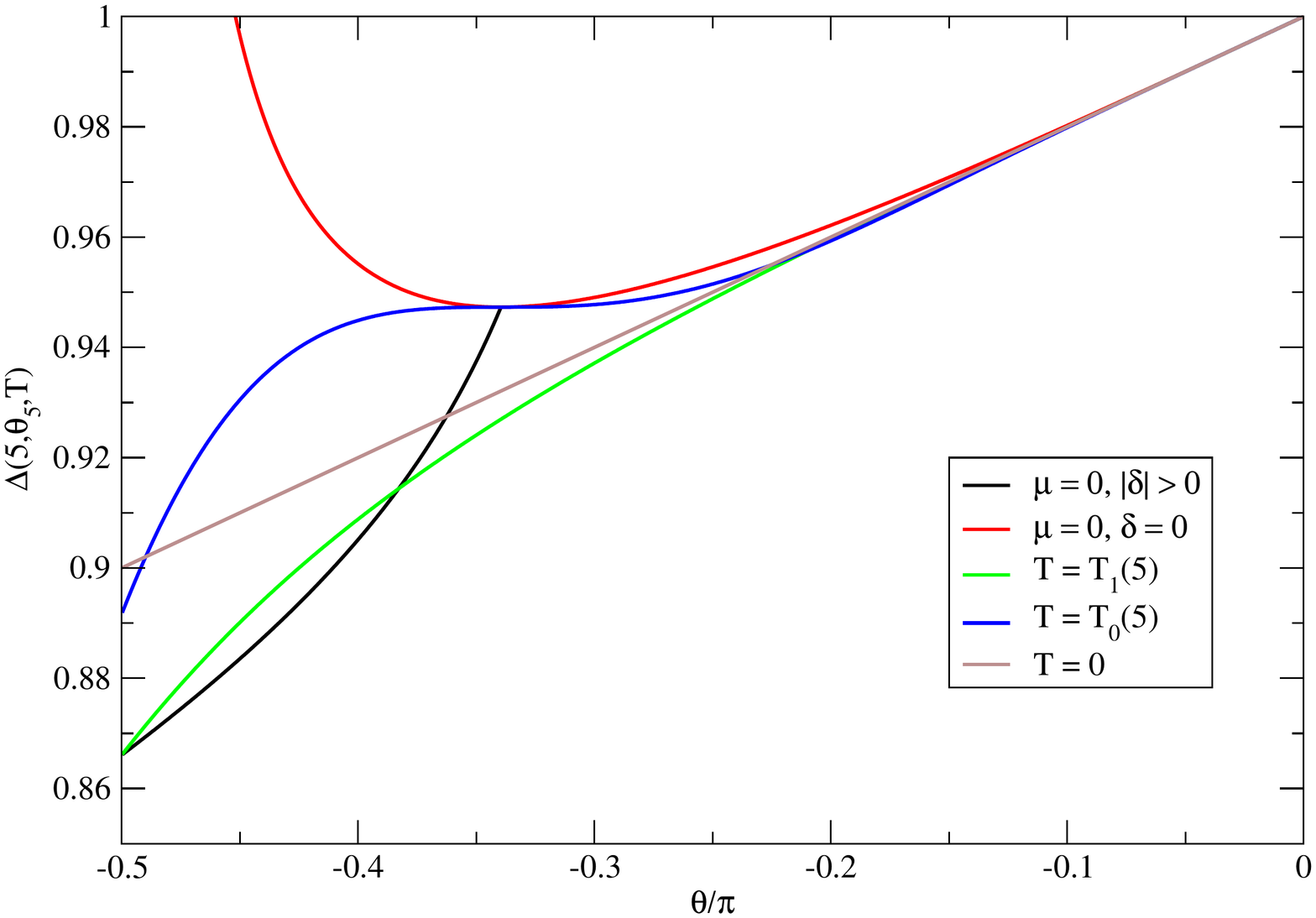}
\caption{ The action at the $n=5$ saddle point is taken as a typical example since it has all the features for a generic $n$ and it is compared with the $n=0$ saddle point to show that the $n=0$ saddle point dominates at all temperatures
and chemical potential. This plot shows the comparison at some specific values of temperature and also at zero chemical potential. }
\eef{bounds5}

After fixing the temperature, we choose a $\theta_n$ in the allowed region and solve for $|u_n|$ using the first equation in \eqn{sadeqn}
and then solve for the chemical potential using the second equation.
Referring to \eqn{derr12ut}, we note that $d_2(|u|,\theta)$ remains non-negative in $-\frac{\pi}{2} \le \theta_n \le 0$ and  we conclude
that $r_2(|u|,\theta)$ is a monotonically increasing function in $-\frac{\pi}{2} \le \theta_n \le 0$. Since we are only considering $\mu > 0$, we can
conclude from the second equation in \eqn{sadeqn} that $u_n > 0$ if $\left[\frac{\pi |u_n|}{T} + r_2(|u_n|,\theta_n)\right]> 0$ and $u_n < 0$ if $\left[\frac{\pi |u_n|}{T}+r_2(|u_n|,\theta_n) \right]< 0$.
Furthermore, we can have a solution to the saddle point equations with $n>0$ and $\mu=0$ only if $\left[ \frac{\pi |u_n|}{T}+r_2(|u_n|,\theta)\right]=0$. 
We have used $n=5$ as a typical value of $n$ and plotted $\Delta\left(5;\theta_5,T_0(5)\right)$ as a blue curve in \fgn{bounds5}. We see it remains positive in the entire range of
$\theta_5$ showing that the saddle at $n=0$ dominates over $n=5$ at $T_0(5)$.

\subsection{$T_1(n)$:}
Consider moving down from $\theta_n=0$. Noting that $u_n$ is positive to start with and noting that
$d_2(|u|,\theta)$ is positive in $-\frac{\pi}{2} \le \theta_n \le 0$, we conclude from \eqn{tddermu} that $\mu$ decreases as
$\theta_n$ moves down from zero as long as $u_n$ remains positive. 
At $\theta_n=-\frac{\pi}{2}$, the first equation in \eqn{sadeqn} reduces to
\be
|u_n| = \frac{(2n-1)\pi}{T}
\ee
and the condition for $\left[ \frac{\pi |u_n|}{T} + r_2\left(|u_n|,-\frac{\pi}{2}\right)\right]$ to remain positive is
\be
\frac{\pi^2}{T^2}(4n^2-1) -\frac{\pi^2}{6} + \sum_{k=1}^\infty \frac{ e^{-\frac{2(2n-1)k\pi}{T}}}{k^2} > 0.
\ee

This gives the condition for $\mu = 0$ to occur for $T > T_n(n)$ with
\bea
T_1(1) = 4.654727000; && T_1(2) = 2\times 4.987847524; \cr
T_1(5) = 5\times 5.044788571; && \lim_{n\to\infty} \frac{T_1(n)}{n} = 5.028967463.
\eea
We have plotted $\Delta\left(5;\theta_5,T_1(5)\right)$ as a red curve in \fgn{bounds5}. We see it remains positive in the entire range of
$\theta_5$ showing that the saddle at $n=0$ dominates over $n=5$ at $T_1(5)$.

\subsection{$T=0$:}
Let us consider the limit $T\to 0$. The two saddle point equations in \eqn{sadeqn} sequentially result in
\be
\lim_{T\to 0} T|u_n| = -\frac{\pi(\theta_n + n\pi)}{\theta_n};\qquad
\lim_{T\to 0} T^2 \left[ \frac{\pi |u_n|}{T} + r_2\left( |u_n| , \theta_n\right) \right] =
\frac{\pi^2(n^2\pi^2-\theta_n^2)}{2\theta^2_n} > 0;
\ee
and we obtain
\be
 \mu =  \frac{\pi(n^2\pi^2-\theta_n^2)}{\theta^2_n};\qquad \delta_n = \frac{\pi(\theta_n + n\pi)^2}{\theta^2_n}.
\ee
With this choice of $\mu$, the saddle point solution with $n=0$ is given by
\be
 \lim_{T\to 0} T u_0  = -\frac{\pi(\theta_n+n\pi)}{\theta_n};\qquad 
\delta_0 = \frac{\pi(\theta_n + n\pi)^2}{\theta^2_n},
\ee
and we conclude
\be
\Delta(n,\theta_n,0) = 
 1+\frac{\theta_n}{n\pi}.
\ee
We have plotted $\Delta(5,\theta_5,0)$ for reference in \fgn{bounds5} as a brown line.

\subsection{$\mu=0$:}

If $T>T_1(n)$ we have a saddle point solution with $\mu=0$ and $n>0$. Setting $\mu=0$, we first note by referring to
\eqn{sadneq0} that the saddle point with $n=0$ corresponds to $u_0=0$ and therefore $\delta_0=0$ and
the real part of the action at the $n=0$ saddle point is (referring to \eqn{actdenri}) 
\be
S_R(0,0,0;0,T) = \frac{2T^2}{\pi} \sum_{k=1}^\infty \frac{(-1)^k}{2k^3} = -\frac{3\zeta(3)}{4\pi} T^2 .
\ee
Referring to \eqn{sadeqn}, the solution to the saddle points at $n>0$ is given by
\be
f(|u_n|,\theta_n)= |u_n| r_1(|u_n|,\theta_n) + (\theta_n + n\pi) r_2(|u_n|,\theta_n)=0.\label{eqnmu0}
\ee
Viewing the above equation as a function of $|u_n|$ at a fixed $\theta_n$ we note that $|u_n|=0$ is a solution to the above equation. 
In this case, $\delta_n=0$ and the temperature as a function of $\theta_n$ from the first equation in \eqn{sadeqn} is
\be
T_n(\theta_n) = \frac{\pi(\theta_n + n\pi)}{r_1(0,\theta_n)},\label{tmu0}
\ee
and
\be
\Delta(n,\theta_n,T_n(\theta_n)) 
= [\frac{\theta_n}{n\pi}+1]^2 + \frac{T_n(\theta_n)}{2n^2\pi^3} J_f(\theta_n);\qquad J_f(\theta) = \sum_{k=1}^\infty (-1)^k \frac{ \cos(2k\theta)-1}{k^3}.\label{actdifmu0}
\ee
Noting that 
\be
J_f\left(-\frac{\pi}{2}\right) =  \frac{7\zeta(3)}{4} > 0;\qquad J_f(0) =0.\label{Jfp}
\ee
and noting that
\be
\frac{dJ_f(\theta)}{d\theta}  = 2 \cl(\pi-2\theta) \le 0,
\ee
where $\cl$ is the Clausen function of order $2$, we conclude that $J_f(\theta)$ is non-negative in $-\frac{\pi}{2} < \theta_n < 0$. Therefore, these saddle points at $n>0$ and $\mu=0$ do not dominate over $n=0$.
We have plotted $\Delta(5,\theta_5,T)$ for $\mu=0$ and $\delta_5=0$ as a red curve in \fgn{bounds5}. Note that the divergence in $\Delta(5,\theta_5,T)$
as $\theta_5\to -\frac{\pi}{2}$ can be seen from \eqn{actdifmu0} and \eqn{Jfp}.

\bef
\centering
\includegraphics[scale=0.45]{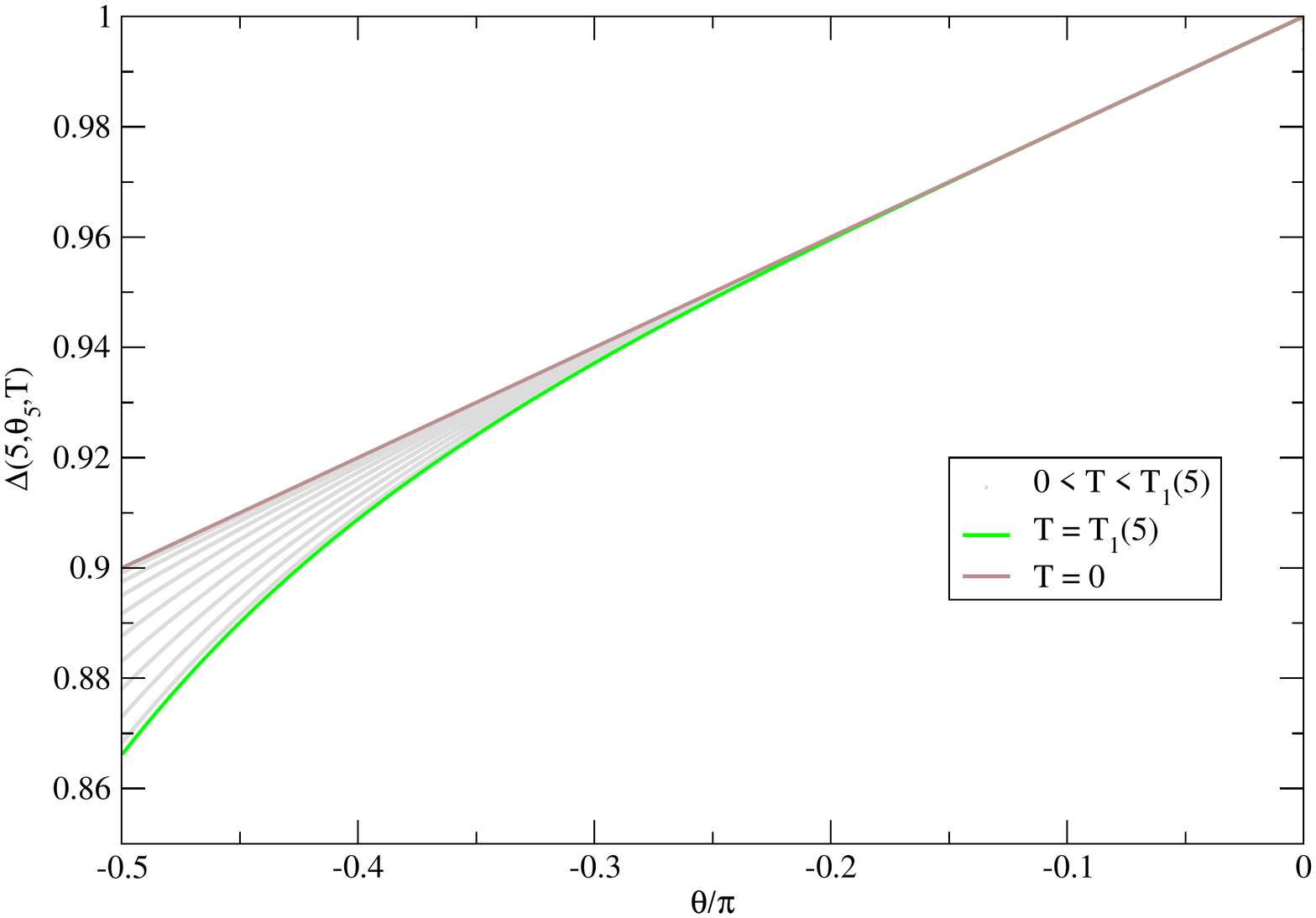}
\includegraphics[scale=0.45]{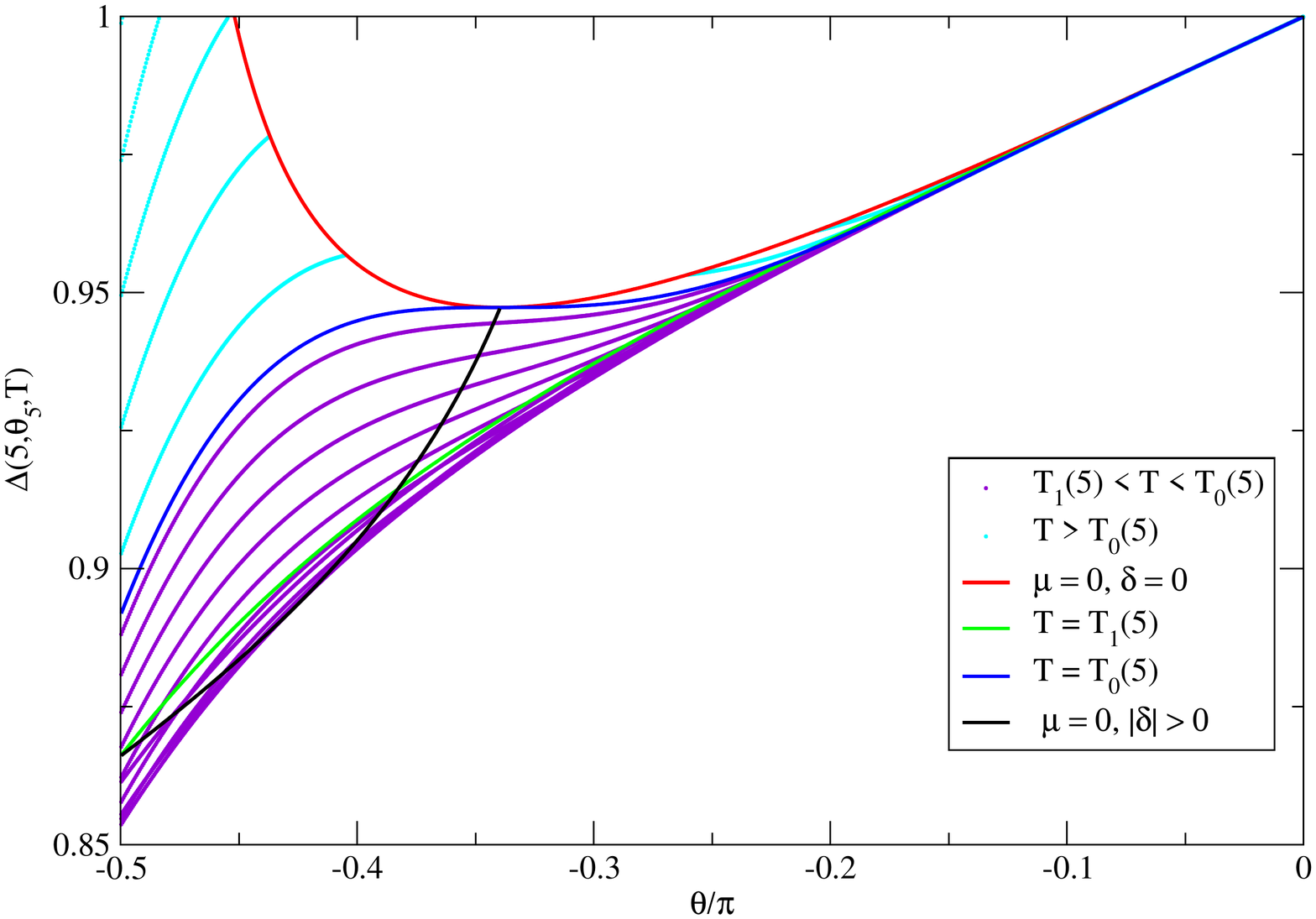}
\caption{ 
The top  plot shows that the difference between $T=0$ and $T=T_1(5)$ is bounded by the two extremes.
The bottom plot shows that the difference between $T=T_1(5)$ and $T=T_0(5)$ moves smoothly from one end to the other but there is
a temperature in between the two ends where the difference is smallest as a function of $\theta_n$. On the other hand,
the difference for $T> T_0(5)$ is bounded from below by $T=T_0(5)$.
}
\eef{n5}

There is another solution to \eqn{eqnmu0} with $|u_n| > 0$ if $ T_1(n) < T < T_0(n)$ and $\theta_n < \theta_0(n)$. To see this note that
the derivative of \eqn{eqnmu0} with respect to $|u_n|$ gives us
\be
\frac{\partial f(|u_n|,\theta_n)}{\partial |u_n|} = r_1(|u_n|,\theta_n) + |u_n| d_2(|u_n|,\theta_n) - (\theta_n + n\pi) d_1(|u_n|,\theta_n).
\ee
The first two terms are positive in the range $-\frac{\pi}{2} < \theta_n < 0$. Furthermore, $d_1(|u_n|,\theta_n)$ is positive only if $-\frac{\pi}{2} < \theta_n < -\frac{\pi}{6}$. We can see that $f(0,\theta_n)=0$ and $f(|u_n|,\theta_n)$ goes as $\left[\frac{n\pi - \theta_n}{2} u_n^2\right]$ as $|u_n|\to \infty$.
If $\frac{\partial f(0,\theta_n)}{\partial |u_n|} > 0$ the only solution to \eqn{eqnmu0} is $|u_n|=0$. The condition for $\frac{\partial f(0,\theta_n)}{\partial |u_n|} = 0$
is the same as \eqn{sadeqnt0} and therefore we conclude that $-\frac{\pi}{2} < \theta_n < \theta_0(n) $ for a solution to \eqn{eqnmu0} to exist with
$|u_n|\ne 0$. We have plotted $\Delta(5,\theta_5,T)$ for $\mu=0$ and $\delta_5\ne 0$ as a black curve in \fgn{bounds5} and it meets the red curve at
$\theta_0(5)$. 
Note that the temperature along the black curve changes from $T=T_1(5)$ at $\theta_5 = -\frac{\pi}{2}$ to $T = T_0(5)$ at
$\theta_5 = \theta_0(5)$. A generic feature of the curves shown in \fgn{bounds5} is the intersection of the black curve ($\mu=0$ and $\delta_5 \ne 0$) with the
green curve ($T_1(5)$) at a value of $\theta_n$ away from $-\frac{\pi}{2}$. The temperature and chemical potential on the two curves at the intersection point are different but yield the same value for $\Delta(n,\theta_n,T)$. 

\subsection{Analysis at $\theta_n=-\frac{\pi}{2}$:}

The two plots in \fgn{n5} shows the behavior of $\Delta(5,\theta_5,T)$ as a function of $\theta_5$ over the entire range of $T$.
For $T < T_1(n)$, we find a solution in the entire range of $\theta_n$ with $\mu$ monotonically increasing from a positive finite value at $\theta_n=-\frac{\pi}{2}$ to
a divergence at $\theta_n=0$. The grey points in the top plot of \fgn{n5}  shows $\Delta(1,\theta_1,T)$ for $T < T_1(n)$ and all points
lie between the line at $T=0$
and $T=T_1(1)$ and moves continuously from one to the other as $T$ changes.
For $T_1(n) < T < T_0(n)$, we again find a solution  in the entire range of $\theta_n$. But we find a $\theta_0(T)$ at which $\mu=0$. The chemical potential
monotonically decreases from a positive finite value at $\theta_n=-\frac{\pi}{2}$ to zero at $\theta_n=\theta_0(T)$ and then monotonically increases to a divergence at $\theta_n=0$. Furthermore, $\theta_0(T_1(n)) = -\frac{\pi}{2}$ and $\theta_0(T_0(n)) = \theta_0(n)$. 
These features are shown in the bottom plot of \fgn{n5}. The curves for $\Delta(5,\theta_5,T)$ at a fixed $T$ in  $T_1(5) < T < T_0(5)$ are
shown by violet points. Note that there is a range of $T$ starting from $T_1(5)$ where the curve intersects the black curve ($\mu=0$ and $\delta_5 \ne 0$) 
at two points. Only one of the intersection points has the same temperature and chemical potential. Initially, it is the intersection point closer to
$\theta_5 = -\frac{\pi}{2}$ till the violet curve is tangential to the black curve. After that, the intersection point closer to $\theta_0(5)$ is where the temperature
and chemical potential matches. 
For $T > T_0(5)$, there will be a solution only if $\theta_5$ does not belong to the interval $\left(\theta_l(T),\theta_r(T)\right)$ which can be obtained
by setting the left hand side to $T$ in \eqn{tmu0}. 
The chemical potential
will monotonically decrease from a positive finite value at $\theta_n=-\frac{\pi}{2}$ to zero at $\theta_l(T)$ and then monotonically increase 
from zero at $\theta_r(T)$ to a divergence at $\theta_n=0$.
These features are also shown in the bottom plot of \fgn{n5} and we see that $\Delta(5,\theta_5,T)$ for $ T > T_0(5)$ shown by cyan points
lies above the curve at $T=T_0(5)$ and has two parts. The boundary of the two parts is the red curve with $\mu=0$ and $\delta=0$.

\bef
\centering
\includegraphics[scale=0.625]{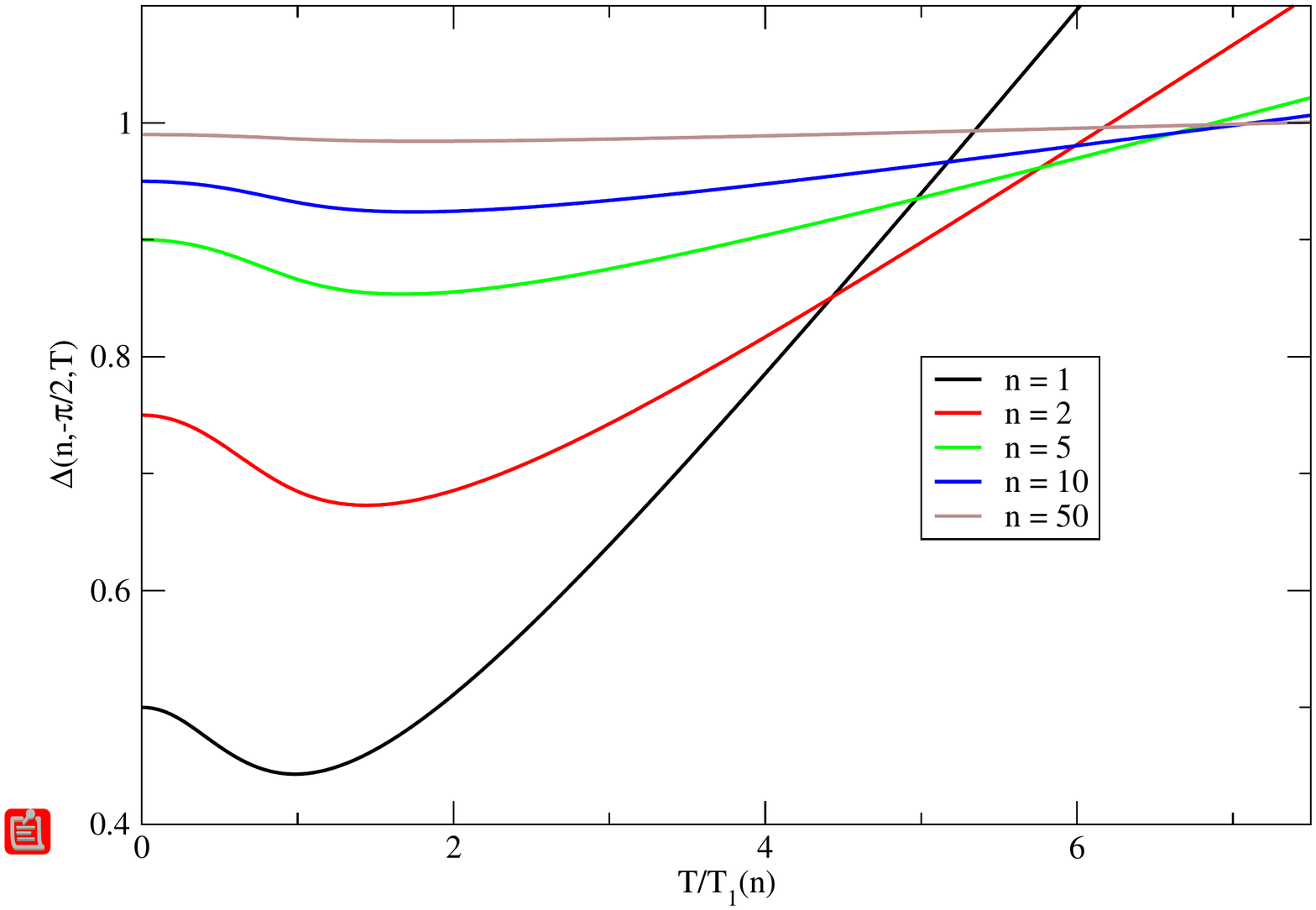}
\caption{ 
The difference in the action between $n>0$ and $n=0$ at $\theta_n=-\frac{\pi}{2}$ is plotted as a function of $T/T_1(n)$.
}
\eef{act-vs-t}
 The graphical analysis discussed before specifically for $n=5$ 
enables us to conclude that $\Delta(5,\theta_5,T)$ reaches a minimum at $\theta_5=-\frac{\pi}{2}$ at any fixed temperature. We therefore plot
$\Delta(n,-\frac{\pi}{2},T)$ as a function of $T$ for $n=1,2,5,10,50$ in \fgn{act-vs-t}. The minimum occurs very close to $T=T_1(1)$ but
moves to a higher temperature with respect to $T_1(n)$ as $n$ is increased. The global minimum occurs at a finite value of $T/T_1(n)$ even as
$n\to\infty$. This plot along with the analysis done specifically for $n=5$ in \fgn{bounds5} and \fgn{n5} serves to graphically prove our main point -- the saddle point at $n=0$ dominates at all chemical potential and temperature.

\section{Conclusions}

We have studied the three dimensional (two spatial and one thermal) Euclidean Thirring model at finite temperature and chemical model in the limit of infinite number of flavors and our aim was to understand the relevance of the Thirring coupling.
The effective action is complex and we had to perform an extensive graphical analysis to show that one particular saddle point dominates over all other possible saddle points at all chemical potential and temperature. The relevance of the Thirring coupling is seen in the dependence of the behavior of the number density as a function of temperature on different values of the number density at zero temperature. The number density as a function of temperature behaves like a three dimensional free fermion gas as the number density at zero temperature approaches zero and it behaves like a two dimensional free fermion gas as the number density at zero temperature approaches infinity. The number density at a fixed non-zero value at zero temperature is a monotonically increasing function of temperature reaching a finite value at infinite temperature. The function for any given non-zero value of density at zero temperature smoothly interpolates between the two extremal behavior at zero numner density and infinite number density at zero temperature.

\begin{acknowledgments}
  The author acknowledges partial support by the NSF under grant
number PHY-1913010.  
\end{acknowledgments}

\appendix

\section{Some useful properties of $r_1(u,\theta)$ and $r_2(u,\theta)$}\label{sec:r1r2rer}

We note that
\be
r_1(|u|,0) =  0;\qquad r_1\left(|u|,\pm\frac{\pi}{2}\right) = \mp \frac{\pi |u|}{2}.\label{ku1}
\ee
Using Riemann and Dirchlet zeta function formulas~\cite{NIST:DLMF}, we also note that
\be
r_2(|u|,0) =  \frac{u^2}{2}  + \int_0^{|u|} dx \ \ln\left(1+e^{-2x}\right);\quad
r_2\left(|u|,\pm\frac{\pi}{2}\right) = \frac{u^2}{2} + \int_0^{|u|} dx \ \ln\left(1-e^{-2x}\right). \label{ku2}
\ee
The derivatives of $r_1(|u|,\theta)$ and $r_2(|u|,\theta)$ with respect to $|u|$ and $\theta$ are
\bea
\frac{\partial r_1(|u|,\theta)}{\partial\theta} = -\frac{\partial r_2(|u|,\theta)}{\partial |u|}
&=& -\frac{1}{2} \ln \left [ 2\left( \cosh(2u)+\cos(2\theta)\right)\right] = d_1(u,\theta).\cr
\frac{\partial r_1(|u|,\theta)}{\partial |u| } = \frac{\partial r_2(|u|,\theta)}{\partial \theta} &=& 
-\tan^{-1} \left(  \tanh |u| \tan\theta \right) 
= d_2(|u|,\theta).\label{derr12ut}
\eea

Referring to the solutions
as $\theta_n(\mu)$, $\delta_n(\mu)$ and $u_n(\mu)$ we compute expressions for the derivative of $\delta_n$ with respect to $\mu$.
Noting that
\be
\frac{ \partial |u_n|}{\partial \mu} = \frac{u_n}{|u_n|} \frac{1}{2T}\left( 1- \frac{\partial \delta_n}{\partial \mu}\right).
\ee
and referring to \eqn{sadeqn}, we obtain
\bea
\frac{\partial \theta_n}{\partial \mu} &=& \frac{ u_n}{|u_n|} \frac{ \pi d_2}{2T^2\left [ \left(\frac{\pi}{T} - d_1(|u_n|,\theta_n)\right)^2 + d_2^2(|u_n|,\theta_n) \right]};\cr
\frac{\partial \delta_n}{\partial \mu} &=&
1 - \frac{ \pi}{\pi - Td_1(|u_n|,\theta_n) + T\frac{d_2^2(|u_n|,\theta_n)}{\frac{\pi}{T} - d_1(|u_n|,\theta_n)}}.\label{tddermu}
\eea

\bibliography{../../mynotes/biblio}

\begin{thebibliography}{29}
\expandafter\ifx\csname natexlab\endcsname\relax\def\natexlab#1{#1}\fi
\expandafter\ifx\csname bibnamefont\endcsname\relax
  \def\bibnamefont#1{#1}\fi
\expandafter\ifx\csname bibfnamefont\endcsname\relax
  \def\bibfnamefont#1{#1}\fi
\expandafter\ifx\csname citenamefont\endcsname\relax
  \def\citenamefont#1{#1}\fi
\expandafter\ifx\csname url\endcsname\relax
  \def\url#1{\texttt{#1}}\fi
\expandafter\ifx\csname urlprefix\endcsname\relax\def\urlprefix{URL }\fi
\providecommand{\bibinfo}[2]{#2}
\providecommand{\eprint}[2][]{\url{#2}}

\bibitem[{\citenamefont{Parisi}(1975)}]{Parisi:1975im}
\bibinfo{author}{\bibfnamefont{G.}~\bibnamefont{Parisi}},
  \bibinfo{journal}{Nucl. Phys.} \textbf{\bibinfo{volume}{B100}},
  \bibinfo{pages}{368} (\bibinfo{year}{1975}).

\bibitem[{\citenamefont{Hikami and Muta}(1977)}]{Hikami:1976at}
\bibinfo{author}{\bibfnamefont{S.}~\bibnamefont{Hikami}} \bibnamefont{and}
  \bibinfo{author}{\bibfnamefont{T.}~\bibnamefont{Muta}},
  \bibinfo{journal}{Prog. Theor. Phys.} \textbf{\bibinfo{volume}{57}},
  \bibinfo{pages}{785} (\bibinfo{year}{1977}).

\bibitem[{\citenamefont{Yang}(1990)}]{Yang:1990ki}
\bibinfo{author}{\bibfnamefont{Z.}~\bibnamefont{Yang}} (\bibinfo{year}{1990}).

\bibitem[{\citenamefont{Gomes et~al.}(1991)\citenamefont{Gomes, Mendes,
  Ribeiro, and da~Silva}}]{Gomes:1990ed}
\bibinfo{author}{\bibfnamefont{M.}~\bibnamefont{Gomes}},
  \bibinfo{author}{\bibfnamefont{R.}~\bibnamefont{Mendes}},
  \bibinfo{author}{\bibfnamefont{R.}~\bibnamefont{Ribeiro}}, \bibnamefont{and}
  \bibinfo{author}{\bibfnamefont{A.}~\bibnamefont{da~Silva}},
  \bibinfo{journal}{Phys. Rev. D} \textbf{\bibinfo{volume}{43}},
  \bibinfo{pages}{3516} (\bibinfo{year}{1991}).

\bibitem[{\citenamefont{Hands}(1995)}]{Hands:1994kb}
\bibinfo{author}{\bibfnamefont{S.}~\bibnamefont{Hands}},
  \bibinfo{journal}{Phys. Rev. D} \textbf{\bibinfo{volume}{51}},
  \bibinfo{pages}{5816} (\bibinfo{year}{1995}), \eprint{hep-th/9411016}.

\bibitem[{\citenamefont{Del~Debbio and Hands}(1996)}]{DelDebbio:1995zc}
\bibinfo{author}{\bibfnamefont{L.}~\bibnamefont{Del~Debbio}} \bibnamefont{and}
  \bibinfo{author}{\bibfnamefont{S.}~\bibnamefont{Hands}},
  \bibinfo{journal}{Phys. Lett. B} \textbf{\bibinfo{volume}{373}},
  \bibinfo{pages}{171} (\bibinfo{year}{1996}), \eprint{hep-lat/9512013}.

\bibitem[{\citenamefont{Hands}(1996)}]{Hands:1996px}
\bibinfo{author}{\bibfnamefont{S.}~\bibnamefont{Hands}}
  (\bibinfo{collaboration}{UKQCD}), in \emph{\bibinfo{booktitle}{{International
  Workshop on Perspectives of Strong Coupling Gauge Theories (SCGT 96)}}}
  (\bibinfo{year}{1996}), pp. \bibinfo{pages}{383--389},
  \eprint{hep-lat/9702003}.

\bibitem[{\citenamefont{Del~Debbio et~al.}(1997)\citenamefont{Del~Debbio,
  Hands, and Mehegan}}]{DelDebbio:1997dv}
\bibinfo{author}{\bibfnamefont{L.}~\bibnamefont{Del~Debbio}},
  \bibinfo{author}{\bibfnamefont{S.}~\bibnamefont{Hands}}, \bibnamefont{and}
  \bibinfo{author}{\bibfnamefont{J.}~\bibnamefont{Mehegan}}
  (\bibinfo{collaboration}{UKQCD}), \bibinfo{journal}{Nucl. Phys. B}
  \textbf{\bibinfo{volume}{502}}, \bibinfo{pages}{269} (\bibinfo{year}{1997}),
  \eprint{hep-lat/9701016}.

\bibitem[{\citenamefont{Del~Debbio and Hands}(1999)}]{DelDebbio:1999he}
\bibinfo{author}{\bibfnamefont{L.}~\bibnamefont{Del~Debbio}} \bibnamefont{and}
  \bibinfo{author}{\bibfnamefont{S.}~\bibnamefont{Hands}},
  \bibinfo{journal}{Nucl. Phys. B} \textbf{\bibinfo{volume}{552}},
  \bibinfo{pages}{339} (\bibinfo{year}{1999}), \eprint{hep-lat/9902014}.

\bibitem[{\citenamefont{Hands and Lucini}(1999)}]{Hands:1999id}
\bibinfo{author}{\bibfnamefont{S.}~\bibnamefont{Hands}} \bibnamefont{and}
  \bibinfo{author}{\bibfnamefont{B.}~\bibnamefont{Lucini}},
  \bibinfo{journal}{Phys. Lett. B} \textbf{\bibinfo{volume}{461}},
  \bibinfo{pages}{263} (\bibinfo{year}{1999}), \eprint{hep-lat/9906008}.

\bibitem[{\citenamefont{Christofi et~al.}(2006)\citenamefont{Christofi, Hands,
  and Strouthos}}]{Christofi:2007ez}
\bibinfo{author}{\bibfnamefont{S.}~\bibnamefont{Christofi}},
  \bibinfo{author}{\bibfnamefont{S.}~\bibnamefont{Hands}}, \bibnamefont{and}
  \bibinfo{author}{\bibfnamefont{C.}~\bibnamefont{Strouthos}}
  (\bibinfo{year}{2006}), \eprint{hep-lat/0703016}.

\bibitem[{\citenamefont{Christofi et~al.}(2007)\citenamefont{Christofi, Hands,
  and Strouthos}}]{Christofi:2007ye}
\bibinfo{author}{\bibfnamefont{S.}~\bibnamefont{Christofi}},
  \bibinfo{author}{\bibfnamefont{S.}~\bibnamefont{Hands}}, \bibnamefont{and}
  \bibinfo{author}{\bibfnamefont{C.}~\bibnamefont{Strouthos}},
  \bibinfo{journal}{Phys. Rev. D} \textbf{\bibinfo{volume}{75}},
  \bibinfo{pages}{101701} (\bibinfo{year}{2007}), \eprint{hep-lat/0701016}.

\bibitem[{\citenamefont{Gies and Janssen}(2010)}]{Gies:2010st}
\bibinfo{author}{\bibfnamefont{H.}~\bibnamefont{Gies}} \bibnamefont{and}
  \bibinfo{author}{\bibfnamefont{L.}~\bibnamefont{Janssen}},
  \bibinfo{journal}{Phys. Rev. D} \textbf{\bibinfo{volume}{82}},
  \bibinfo{pages}{085018} (\bibinfo{year}{2010}), \eprint{1006.3747}.

\bibitem[{\citenamefont{Janssen and Gies}(2012)}]{Janssen:2012pq}
\bibinfo{author}{\bibfnamefont{L.}~\bibnamefont{Janssen}} \bibnamefont{and}
  \bibinfo{author}{\bibfnamefont{H.}~\bibnamefont{Gies}},
  \bibinfo{journal}{Phys. Rev. D} \textbf{\bibinfo{volume}{86}},
  \bibinfo{pages}{105007} (\bibinfo{year}{2012}), \eprint{1208.3327}.

\bibitem[{\citenamefont{Schmidt et~al.}(2016)\citenamefont{Schmidt,
  Wellegehausen, and Wipf}}]{Schmidt:2015fps}
\bibinfo{author}{\bibfnamefont{D.}~\bibnamefont{Schmidt}},
  \bibinfo{author}{\bibfnamefont{B.}~\bibnamefont{Wellegehausen}},
  \bibnamefont{and} \bibinfo{author}{\bibfnamefont{A.}~\bibnamefont{Wipf}},
  \bibinfo{journal}{PoS} \textbf{\bibinfo{volume}{LATTICE2015}},
  \bibinfo{pages}{050} (\bibinfo{year}{2016}), \eprint{1511.00522}.

\bibitem[{\citenamefont{Wellegehausen et~al.}(2017)\citenamefont{Wellegehausen,
  Schmidt, and Wipf}}]{Wellegehausen:2017goy}
\bibinfo{author}{\bibfnamefont{B.~H.} \bibnamefont{Wellegehausen}},
  \bibinfo{author}{\bibfnamefont{D.}~\bibnamefont{Schmidt}}, \bibnamefont{and}
  \bibinfo{author}{\bibfnamefont{A.}~\bibnamefont{Wipf}},
  \bibinfo{journal}{Phys. Rev. D} \textbf{\bibinfo{volume}{96}},
  \bibinfo{pages}{094504} (\bibinfo{year}{2017}), \eprint{1708.01160}.

\bibitem[{\citenamefont{Hands}(2017)}]{Hands:2017hhk}
\bibinfo{author}{\bibfnamefont{S.}~\bibnamefont{Hands}}, in
  \emph{\bibinfo{booktitle}{{35th International Symposium on Lattice Field
  Theory}}} (\bibinfo{year}{2017}), \eprint{1708.07686}.

\bibitem[{\citenamefont{Hands}(2019)}]{Hands:2018vrd}
\bibinfo{author}{\bibfnamefont{S.}~\bibnamefont{Hands}},
  \bibinfo{journal}{Phys. Rev. D} \textbf{\bibinfo{volume}{99}},
  \bibinfo{pages}{034504} (\bibinfo{year}{2019}), \eprint{1811.04818}.

\bibitem[{\citenamefont{Hands}(2018)}]{Hands:2018kvr}
\bibinfo{author}{\bibfnamefont{S.}~\bibnamefont{Hands}}, \bibinfo{journal}{PoS}
  \textbf{\bibinfo{volume}{Confinement2018}}, \bibinfo{pages}{221}
  (\bibinfo{year}{2018}).

\bibitem[{\citenamefont{Lenz et~al.}(2019)\citenamefont{Lenz, Wellegehausen,
  and Wipf}}]{Lenz:2019qwu}
\bibinfo{author}{\bibfnamefont{J.~J.} \bibnamefont{Lenz}},
  \bibinfo{author}{\bibfnamefont{B.~H.} \bibnamefont{Wellegehausen}},
  \bibnamefont{and} \bibinfo{author}{\bibfnamefont{A.}~\bibnamefont{Wipf}},
  \bibinfo{journal}{Phys. Rev. D} \textbf{\bibinfo{volume}{100}},
  \bibinfo{pages}{054501} (\bibinfo{year}{2019}), \eprint{1905.00137}.

\bibitem[{\citenamefont{Karthik and
  Narayanan}(2016{\natexlab{a}})}]{Karthik:2015sgq}
\bibinfo{author}{\bibfnamefont{N.}~\bibnamefont{Karthik}} \bibnamefont{and}
  \bibinfo{author}{\bibfnamefont{R.}~\bibnamefont{Narayanan}},
  \bibinfo{journal}{Phys. Rev.} \textbf{\bibinfo{volume}{D93}},
  \bibinfo{pages}{045020} (\bibinfo{year}{2016}{\natexlab{a}}),
  \eprint{1512.02993}.

\bibitem[{\citenamefont{Karthik and
  Narayanan}(2016{\natexlab{b}})}]{Karthik:2016ppr}
\bibinfo{author}{\bibfnamefont{N.}~\bibnamefont{Karthik}} \bibnamefont{and}
  \bibinfo{author}{\bibfnamefont{R.}~\bibnamefont{Narayanan}},
  \bibinfo{journal}{Phys. Rev.} \textbf{\bibinfo{volume}{D94}},
  \bibinfo{pages}{065026} (\bibinfo{year}{2016}{\natexlab{b}}),
  \eprint{1606.04109}.

\bibitem[{\citenamefont{Karthik and Narayanan}(2017)}]{Karthik:2017hol}
\bibinfo{author}{\bibfnamefont{N.}~\bibnamefont{Karthik}} \bibnamefont{and}
  \bibinfo{author}{\bibfnamefont{R.}~\bibnamefont{Narayanan}},
  \bibinfo{journal}{Phys. Rev.} \textbf{\bibinfo{volume}{D96}},
  \bibinfo{pages}{054509} (\bibinfo{year}{2017}), \eprint{1705.11143}.

\bibitem[{\citenamefont{Karthik and
  Narayanan}(2016{\natexlab{c}})}]{Karthik:2016ixj}
\bibinfo{author}{\bibfnamefont{N.}~\bibnamefont{Karthik}} \bibnamefont{and}
  \bibinfo{author}{\bibfnamefont{R.}~\bibnamefont{Narayanan}},
  \bibinfo{journal}{PoS} \textbf{\bibinfo{volume}{LATTICE2016}},
  \bibinfo{pages}{245} (\bibinfo{year}{2016}{\natexlab{c}}),
  \eprint{1610.09355}.

\bibitem[{\citenamefont{Karthik and Narayanan}(2018)}]{Karthik:2018tnh}
\bibinfo{author}{\bibfnamefont{N.}~\bibnamefont{Karthik}} \bibnamefont{and}
  \bibinfo{author}{\bibfnamefont{R.}~\bibnamefont{Narayanan}},
  \bibinfo{journal}{Phys. Rev. Lett.} \textbf{\bibinfo{volume}{121}},
  \bibinfo{pages}{041602} (\bibinfo{year}{2018}), \eprint{1803.03596}.

\bibitem[{\citenamefont{Karthik and Narayanan}(2019)}]{Karthik:2019mrr}
\bibinfo{author}{\bibfnamefont{N.}~\bibnamefont{Karthik}} \bibnamefont{and}
  \bibinfo{author}{\bibfnamefont{R.}~\bibnamefont{Narayanan}}
  (\bibinfo{year}{2019}), \eprint{1908.05500}.

\bibitem[{\citenamefont{Goykhman}(2016)}]{Goykhman:2016zgd}
\bibinfo{author}{\bibfnamefont{M.}~\bibnamefont{Goykhman}},
  \bibinfo{journal}{JHEP} \textbf{\bibinfo{volume}{07}}, \bibinfo{pages}{034}
  (\bibinfo{year}{2016}), \eprint{1605.08449}.

\bibitem[{\citenamefont{Gradshteyn and Ryzhik}(2007)}]{GradRyz:2007}
\bibinfo{author}{\bibfnamefont{I.}~\bibnamefont{Gradshteyn}} \bibnamefont{and}
  \bibinfo{author}{\bibfnamefont{I.}~\bibnamefont{Ryzhik}},
  \bibinfo{journal}{Academic Press}  (\bibinfo{year}{2007}).

\bibitem[{{\relax DLMF}()}]{NIST:DLMF}
{\relax DLMF}, \emph{\bibinfo{title}{{\it NIST Digital Library of Mathematical
  Functions}}}, \bibinfo{howpublished}{http://dlmf.nist.gov/, Release 1.0.26 of
  2020-03-15}, \bibinfo{note}{f.~W.~J. Olver, A.~B. {Olde Daalhuis}, D.~W.
  Lozier, B.~I. Schneider, R.~F. Boisvert, C.~W. Clark, B.~R. Miller, B.~V.
  Saunders, H.~S. Cohl, and M.~A. McClain, eds.},
  \urlprefix\url{http://dlmf.nist.gov/}.

\end{thebibliography}
\end{document}